\documentclass[12pt]{article}
\usepackage{amssymb}
\usepackage{graphicx,enumerate}

\newcommand {\beq} {\begin{equation}}
\newcommand {\eeq} {\end{equation}}

\begin{document}
\title{Impossibility Proofs and Quantum Bit Commitment}
\author{Horace P.~Yuen\footnote{yuen@eecs.northwestern.edu}\hspace{2mm} \\
% Center for Photonic Communication and Computing\\
Department of Electrical Engineering and Computer Science\\
Department of Physics and Astronomy\\
Northwestern University, Evanston, IL 60208} \maketitle
\begin{abstract} The nature and scope of various impossibility proofs as they relate to  real-world situations are discussed. In particular, it is shown in words without technical symbols how secure quantum bit commitment protocols may be obtained with testing that exploits the multiple possibilities of cheating entanglement formation.
 \end{abstract}

\section{Introduction}

In this paper I would like to discuss the general issue of the scope of ``impossibility proof'' of various different kinds, ranging from the classical straightedge/compass construction to hidden-variables and bit commitment in quantum physics. The aim is to highlight the difficulties in characterizing all the possibilities that can be obtained in principle in the real world, which are purportedly ruled out by the impossibility proofs. The issue is of fundamental significance for understanding the relation between real world features and their mathematical representations, a subject of great importance and subtlety in general. More specifically, I would also point out the non-existence of a universal quantum bit commitment (QBC) impossibility proof. The presumed existence of such a proof is widely held, which sociologically and psychologically closes out a field that is rare in the area of quantum information in its potentially significant and realistic applications. A new way for obtaining secure QBC protocols would also be indicated.

\section{Five Cases of Impossibility Claims}

Consider the following list of impossibility (non-existence) claims, in chronological order of their first appearance, which are supported by ``proofs'' of various sorts to be analyzed.

\begin{enumerate} [(i)]
\item No Trisection of an Arbitrary Angle with Straight Edge and Compass: \\ \\
It is not possible to construct in a finite number of steps an angle equal to one-third of an arbitrarily given angle using only a compass and an unmarked straight edge.

\item Church-Turing Thesis: \\ \\
It is not possible to find a mechanical procedure that cannot be implemented by a Turing machine.

\item No-Hidden Variable Theorem of von Neumann: \\ \\
There is no hidden variable theory that would reproduce the predictions of quantum mechanics.

\item No-Clone Theorem: \\ \\
There is no physical system that would produce at the output two identical copies of an input quantum state drawn from a set of two nonorthogonal states.

\item Impossibility of Quantum Bit Commitment: \\ \\
There is no QBC protocol that is arbitrarily close to being perfectly secure for both parties.

\end{enumerate}

What is remarkable about an impossibility theorem of the above variety is that it rules out not just mathematical possibilities but physical realistic ones. This implies that all the relevant physical possibilities in the real world for the problem situation under consideration have been taken into account in the mathematical formulation. Even given a physical theory describing the real world such as quantum physics, this may or may not be possible depending on whether one can characterize mathematically, or at least include in a mathematical description, \emph{all} the apparent possibilities. This is because generally we know of no mathematical characterization of all the possible real world referents of a given clearly meaningful natural language sentence. We illustrate this point with (i), perhaps the oldest impossibility claim.

How do we characterize all possible straight edge/compass constructions so we may prove, say, it is a logical contradiction if any one of them can be used to trisect an arbitrary angle. Indeed, from the standard impossibility proof as obtained from Galois theory, that is not possible even for some specific angles $\theta$, say $\theta=\pi/6$. On the other hand, it is well-known that such construction is possible if the straight edge is marked, or if $\cdots$, and so on. A simple such construction is given by Archimedes \cite{bold69}, the greatest ancient Greek mathematician according to many, which works as illustrated in Fig.~1. How does one exclude it in the description ``compass and unmarked straight edge''?

Given a compass, I contend that there is no need to mark a straight edge in order to get an effectively marked one as follows. One sets the basic unit measure with the compass, e.g., from $AB$ of Fig.~1, and then flips it along the straight edge to get any integer multiple. For the Archimedean construction, one could slide the compass pointer along the straight line $AC$ and see when the pencil would cut the circle. I do not wish to go into Zeno's paradoxes here, but the above seems to be a real world construction one can carry out with a compass and straight edge.

Of course such operations are not allowed in the intended description of ``compass and unmarked straight edge''. A much less misleading description of the allowed operations can be given in this case with a more precise specification --- indeed a perfectly precise one is given algebraically which is to be translated to the ``compass and straight edge'' language in some way. My point here is that the algebraic restriction is precise but the corresponding ``compass and straight edge'' description is not, certainly not for their possible operations in the real world. There are manifold problems for such precise translation, which I think lie at the foundation of much human knowledge with little known results.

My conclusion on this case (i) is that there is no mathematical characterization of all possible ``compass and unmarked straight edge'' operations in the real world. The impossibility proof works for a very specific subset of such operations which are taken, perhaps appropriately when described in sufficient detail, to correspond exactly to a set of algebraic operations.

The significance and difficulty of characterizing all possible real world operations, even just in classical physics, is illustrated by case (ii). It is taken to be an empirical fact that a Turing machine, or any of its equivalents such as a Post machine, can simulate any mechanical procedure. The claim is rightly called a thesis, \emph{not} a theorem, not only because it is unproven, but because one cannot have a theorem on something in the real world -- in this case, a mechanical procedure -- which does not correspond to a primitive of the mathematical system and also has no mathematical definition in terms of the primitives.

This lack of a mathematical definition for ``mechanical procedure'' does not prevent the thesis from being meaningful in the real world as long as one can recognize some, maybe a lot -- but all is not required -- procedures to be mechanical. However, it may be remarked that Zeno's paradoxes and corresponding ``infinite machines'' \cite{salmon70} can be construed as contrary claims to the Church-Turing thesis, precisely because ``mechanical procedure'' is not a precisely characterized concept.

Case (iii) on von Neumann's No-Hidden Variable Theorem \cite{bell66} provides a clear illustration of the problem of how one may characterize mathematically all systems or processes of a certain kind, in this case a ``hidden-variable theory''. I would not discuss here the now well known suspension of von Neumann's fifth linearity requirement of a hidden variable theory, which is deemed unnecessary for both local and nonlocal such theories. A deeper problem still exists on what one means by a ``local theory'', which is exemplified by the recent controversy on the non-existence of local hidden-variable theory supposedly given by the ``Bell Theorem'' \cite{christian07}.

The No-Clone Theorem of case (iv) is indeed a theorem concerning all processes described by quantum physics, because all such processes can be included in a proper mathematical representation. The usual proof \cite{yuen86} contains the basic ingredients of a complete proof that can be spelled out readily.

\section{Quantum Bit Commitment}

For the rest of this article I will talk about case (v) on QBC. \emph{At best} a general QBC impossibility claim would have the status of a thesis as in case (ii) for Turing machines as Ozawa emphasized, for the similar reason that no one knows the mathematical characterization of all QBC protocols, or indeed of all protocols of any kind, classical or quantum. As in the case of a mechanical procedure, one can typically recognize a QBC protocol when presented one although one cannot include them all in a mathematical formulation.

In this connection, it may be observed that it is possible to characterize all attacks on a given protocol of any kind, although that is already a somewhat subtle issue -- the relation between reality and its mathematical representation is almost always a subtle issue from the vantage point of present human knowledge. Note also that if there is a problem of characterizing all attacks on a protocol, it would not be possible to give unconditional security proofs such as those claimed for quantum key distribution.

Despite this, it is widely accepted that there is a universal impossibility \emph{proof} on secure QBC since 1997 \cite{mayers97,lochau97}, and the recent paper \cite{dariano07} adds to this impression despite its ambivalence on this point. The implicit claim has always been that \emph{all} possible QBC protocols have been included in the mathematical formulation of the problem given in the paper. The fact that new features were introduced that were not covered in previous formulations has always been ignored, as long as the feature has not led to a widely accepted secure protocol which is the case thus far. The general impossibility claim is repeated with the insistence that there is a \emph{proof} for it.

There have been contrary claims from time to time appearing in the quant-ph archive that secure QBC is possible with specific protocols and security proof sketches. Unfortunately, it is notoriously difficult to understand someone else's QBC protocol and security proof. Thus, I confine myself in the following to only the \emph{new} features I myself introduced in protocols of various sorts that have not been covered in \emph{previous} impossibility proofs. The discussion would be just in words and hopefully understandable to anyone somewhat knowledgeable on the issue as in cases (i)-(iv) above. The intention is to lay out the issues in a way that preliminary assessment can be made without going into technical details.

It is important to re-emphasize that there are \emph{two distinct issues} here. One is whether there is a universal impossibility proof and the other is whether there is a QBC protocol that has been proved secure. The very fact that some features were not covered in previous proofs shows that there was \emph{no valid basis} to claim the existence of a universal impossibility proof, regardless of whether any secure protocol has been found.
In the following, I will discuss a specific new feature not covered in any of the impossibility proofs and which leads to secure protocols.

Broadly speaking, I have given three different new ways to obtain a possibly secure protocol since 2000 at the QCMC in Capri, Italy. At least two of these were not covered by any impossibility proof known to anyone at the time they were first discussed, in print or in saying. They are:-
\begin{enumerate}[(1)]
\item Use of Anonymous States: \\
A party uses classical random numbers in the protocol not known to the other party.
%\newpage
\item Action from B to Prevent Entanglement Cheating from A: \\
The idea is for B to destroy A's (cheating) entanglement by A's action.

\item Testing before Commitment or Opening: \\
The idea is to destroy A or B's own (cheating) entanglement by demanding answers which force measurement on his/her own ancilla.

\end{enumerate}

In case (1) with anonymous states, much initial reaction was that the classical randomness can be purified quantum mechanically and the resulting pure state is assumed openly known. This is clearly not a realistic assumption, and does not obtain under what I call the Secrecy Principle \cite{yuen03qcmc} for QBC protocols: Whatever a party does is not required to be revealed to the other party if it does not permit the first party to cheat. Clearly, such use of classical random numbers must be allowed as it is in QKD protocols. It has nothing to do with Kerckhoffs' principle as alleged in \cite{dariano07}. In this connection, it may be noted that something must be kept private in any kind of secure protocol. In the QBC case, it is not realistic to assume that a party knows the randomness purification basis of the other party. A theorem based on such an assumption has no real world application.

It turns out it appears impossible to get a secure protocol this way. For perfect concealing, a general proof of this impossibility was given in \cite{yuen01} for a two-round protocol. A different argument applicable to multi-round protocols was given by Ozawa \cite{ozawa01} and later by Cheung \cite{cheung05}. For approximate concealing, a simple proof covering all natural protocols was given by me in a slide prepared for the 2005 QIT meeting in Sendai, Japan. More complicated versions are also given in \cite{dariano07} and \cite{cheung06}.

In situation (2), the general futility of B's action follows from the simple commutativity of A and B's actions when there is no checking intervention in between. While it can reasonably be maintained that such possibility is implicitly covered by the known impossibility proofs due to its technical simplicity, it is conceptually far from trivial and should be indicated in a general impossibility proof, especially one that goes not by deriving a general contradiction but by examining each party's possible actions.

In this connection, it may be observed there was really no place where one can find a systematic QBC impossibility proof in the literature other than vague generalities, till the appearance of \cite{dariano07}.  This is indicative of the state of the field. Unfortunately, the formulation and proof in \cite{dariano07} is not given or translated into the usual language, and thus is very difficult to understand and to assess. It is not even clear whether a universal impossibility proof is claimed.

The testing possibilities for case (3) are clearly legitimate protocol elements -- similar ones are used in QKD protocols. They were simply not covered in the impossibility proofs, not completely even in \cite{dariano07}. It was found that entanglement cheating may be retained for some such proposed protocols via proper L$\ddot{u}$ders measurement on part of the ancilla. Indeed, it turns out that if a party has the choice of many possible entanglements that yield the same mixed result for the other party, such as a full permutation entanglement for an unknown bit position, there is no known way to create a secure protocol this way. On the other hand, if the party is limited to use a \emph{specific entanglement}, it can be destroyed by his/her own measurement upon answering truthfully to questions. This possibility is explicitly described in \cite{yuen07} as my protocol QBC3. It works in a similar way in a different setting as my protocol QBC1 \cite{yuen03}.

How does one party know a specific entanglement has actually been carried out instead of some other action by the other party? In the QBC literature, the claim has always been that even under \emph{honest entanglement} with all classical randomness purified, no protocol can be secure \cite{brassard97,spekkens01}. Such claim is made without allowance for checking. There was no discussion on even what would happen if one is caught cheating during protocol execution. The usual implicit assumption is that there is no need for anyone to cheat during testing. Nevertheless, it is important to note that whether a party is using the prescribed entanglement \emph{can} indeed be checked before commitment in many protocols including my QBC1 and QBC3. In contrast, this cannot be accomplished in a single-pass protocol to prevent A's entanglement cheating.

Intuitively, a party's entanglement cheating possibility may be destroyed when he/she has to measure on his/her ancilla to provide correct answers during testing. If the entanglement is sparse, the measurement would totally disentangle the corresponding bit sequence but it \emph{still appears mixed} to the other party in the form of classical randomness with no quantum purification. The situation is similar to the case of one-pass protocols in which A picks a state randomly without actual entanglement. The protocol is concealing but A cannot cheat anyway. On the other hand, there can be no checking in one-pass protocols so A cannot be assumed to be ``honest'' and thus not entangle the choices.

This subtle possibility in my QBC1 and QBC3 was overlooked. The multiple possible entanglements all work the same if there is no testing later. In fact, the one with minimal ancilla dimension is suggested in \cite{dariano07}. Thus, what is missed in \cite{dariano07} is the possibility that a specific entanglement is \emph{dictated} by the protocol which upon testing can be destroyed to prevent further cheating. The situation here is in a way similar to case (i) on angle trisection in the difficulty of covering all real world possibilities. It is similar to case (iii) on no hidden-variables in that only a specific formulation is covered that does not exhaust all possible QBC protocols one can construct.

It may be pointed out that one cannot dismiss the above secure possibility by claiming there is no (approximate) concealing proof for protocols QBC1 and QBC3. There is. But even if there is none, the possibility is not covered by known impossibility proofs because such multiple entanglement possibility for the same classical randomness is not or has not been related to concealing. Here is an example of the two distinct issues mentioned above.

\section{Conclusion}

Secure bit commmitment is an extraordinarily useful tool that may be used to perform many tasks. The area of quantum information would be enormously enriched if QBC is included in its development. Note that the entanglement cheating needed against a QBC protocol is not even on the experimental horizon and may never be. Indeed, many QBC protocols are more \emph{practically secure} than any known QKD protocol and  proper quantitative security comparison between them deserves to be made. That fully secure QBC protocols cannot be obtained as claimed by the QBC impossibility proofs constitutes a sociological and psychological barrier to the development of QBC. It is a difficult task to overcome this widely held perception, which I hope to begin with this paper. One should be able to tell, independently of my own specific protocols, that secure quantum bit commitment is far from a settled issue.

\section{Acknowledgement}\nonumber

I would like to thank C.Y.~Cheung, G.M.~D'Ariano, R.~Nair and M.~Ozawa for many useful discussions.

\section{Archimedes's Trisection of an Arbitrary Angle}

\begin{figure*}
\vspace*{-6pt}
\centerline{\includegraphics[width=0.7\textwidth]{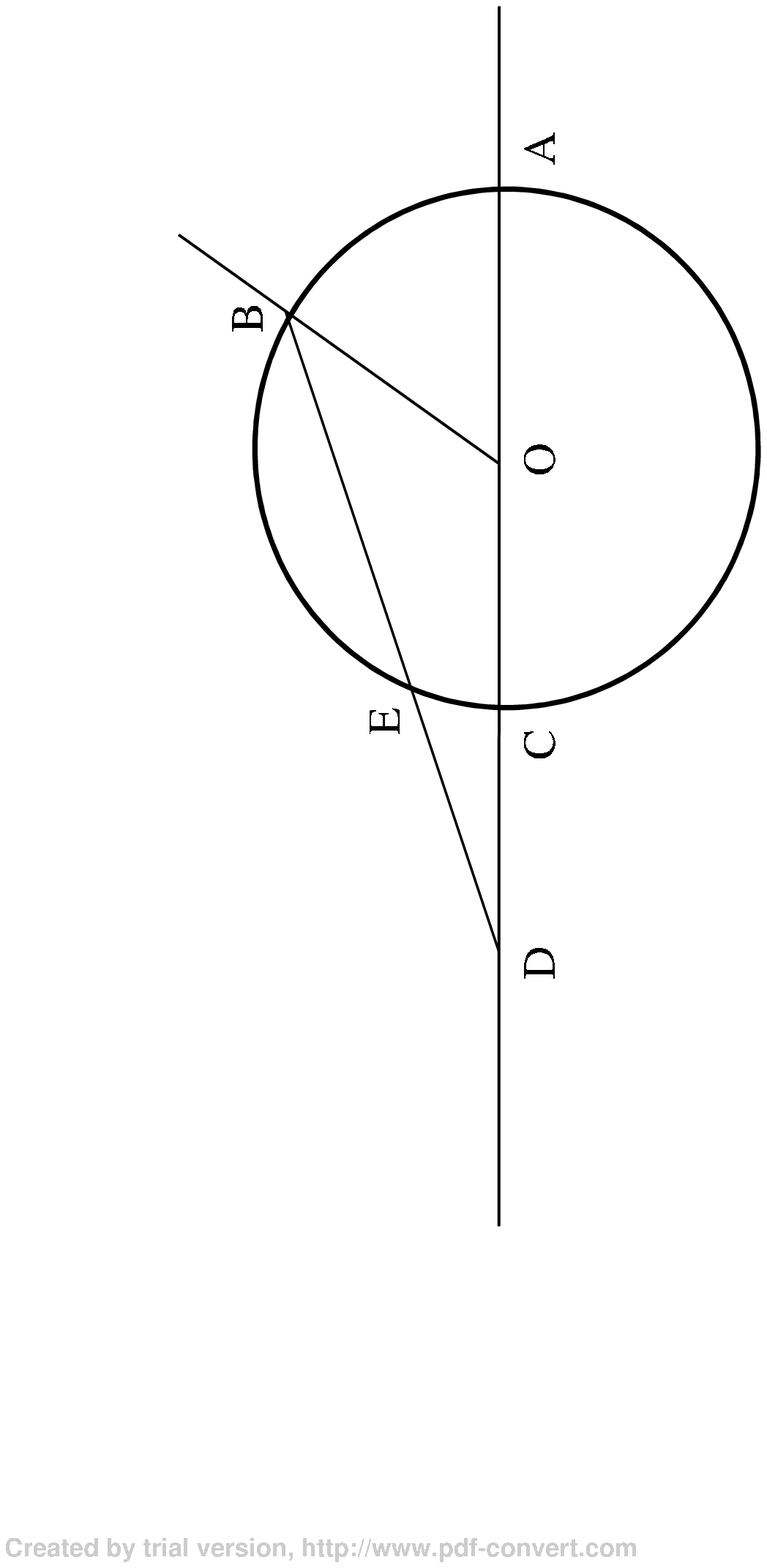}}
\caption{Archimedes's trisection of an arbitrary angle}
\end{figure*}

In Fig.~1, let $AOB$ be a given angle. Construct a circle with $O$ as center and any length as radius. Construct a line through $B$ intersecting diameter $AC$ extended so that $ED$ is equal to the radius $AO$. The angle $CDE$ is one third of angle $AOB$.

\end{document}